%% file: main.tex
\documentclass[10pt,journal,compsoc]{IEEEtran}
\ifCLASSOPTIONcompsoc
  \usepackage[nocompress]{cite}
\else
  \usepackage{cite}
\fi
\ifCLASSINFOpdf
\else
\fi
\usepackage{amsmath,amssymb,amsfonts}
\usepackage{algpseudocode}
\usepackage{algorithm}
\usepackage{pifont}
\usepackage{graphicx}
\usepackage{textcomp}
\usepackage{xcolor}
\usepackage{pgf}
\usepackage{tikz}
\usepackage{color}
\usepackage{stfloats}
\usepackage{multirow}
\usepackage{diagbox}
\usepackage[hyphens]{url}
\input{extra_command}

\begin{document}
%
\title{
Accelerating GNN Training through Locality-aware Dropout and Merge
}
%
%
%
%

\author{Gongjian~Sun,
        Mingyu~Yan,~\IEEEmembership{Member,~IEEE},
        Dengke~Han, \\
        Runzhen~Xue,
        Duo~Wang,
        Xiaochun~Ye, 
        and~Dongrui~Fan~\IEEEmembership{Senior~Member,~IEEE}
\IEEEcompsocitemizethanks{
\IEEEcompsocthanksitem 
Gongjian Sun, Mingyu Yan, Dengke Han, Runzhen Xue, Duo Wang, Xiaochun Ye, Dongrui Fan are with the State Key Lab of Processors, Institute of Computing Technology, Chinese Academy of Sciences, Beijing, China, and also with the University of Chinese Academy of Sciences, Beijing, China
(e-mail: sungongjian15@mails.ucas.ac.cn, \{yanmingyu, handengke21s, xuerunzhen21s, wangduo, yexiaochun, fandr\}@ict.ac.cn). 

\IEEEcompsocthanksitem 
Mingyu Yan is the corresponding author.
}


}

%
%

\markboth{IEEE Transactions on Parallel and Distributed Systems}
{\todo{who} \MakeLowercase{\textit{et al.}}: \todo{title}}
%



\IEEEtitleabstractindextext{%
\input{tex/abstract}

}

\maketitle

\IEEEdisplaynontitleabstractindextext

%
\IEEEpeerreviewmaketitle

\input{tex/intro}
\input{tex/background}
\input{tex/motivation}

\input{tex/design}
\input{tex/evaluation}
\input{tex/related_work}
\input{tex/conclusion}



\ifCLASSOPTIONcompsoc
    \section*{Acknowledgments}
\else
    \section*{Acknowledgment}
\fi

This work was supported by National Key Research and Development Program (Grant No. 2022YFB4501400), the National Natural Science Foundation of China (Grant No. 62202451 and No. 62302477), CAS Project for Young Scientists in Basic Research (Grant No. YSBR-029), and CAS Project for Youth Innovation Promotion Association.


\ifCLASSOPTIONcaptionsoff
  \newpage
\fi


\bibliographystyle{IEEEtranS}
\bibliography{references}

%

\input{tex/bio}










\end{document}

%% file: extra_command.tex
\usepackage{soul}
\usepackage{color}
\usepackage{multirow}
\usepackage{makecell}
\usepackage{xcolor}
\usepackage{amssymb}
\usepackage{tikz}
\usetikzlibrary{calc}
\usepackage[bookmarks=true,breaklinks=true,colorlinks,linkcolor=blue,citecolor=blue,urlcolor=black]{hyperref}

\usepackage{booktabs}
\usepackage{comment}
\usepackage{ulem}

\newcommand{\squishlist}{
   \begin{list}{$\bullet$}
    { \setlength{\itemsep}{0pt}      \setlength{\parsep}{0pt}
      \setlength{\topsep}{3pt}       \setlength{\partopsep}{0pt}
      \setlength{\listparindent}{-2pt}
      \setlength{\itemindent}{-5pt}
      \setlength{\leftmargin}{1em} \setlength{\labelwidth}{0em}
      \setlength{\labelsep}{0.5em} } }

\newcommand{\squishend}{
    \end{list}  }

\newcommand{\todo}[1]{{\color{red}\sf\bfseries \hl{[TODO]} #1}}

\setuldepth{cat}

\newcommand{\note}[1]{{\color{magenta}$\square$}}

\newcounter{Emp}[subsection]

\newcounter{TimeEmp}

\newcounter{ComputeEmp}

\newcounter{MemoryEmp}

\newcounter{CommunicationEmp}

\newcounter{ParallelismEmp}

\newcounter{DataflowEmp}

\newcounter{ExecutionEmp}

%% file: tex/abstract.tex
\begin{abstract}
Graph Neural Networks (GNNs) have demonstrated significant success in graph learning and are widely adopted across various critical domains. However, the irregular connectivity between vertices leads to inefficient neighbor aggregation, resulting in substantial irregular and coarse-grained DRAM accesses. This lack of data locality presents significant challenges for execution platforms, ultimately degrading performance. While previous accelerator designs have leveraged on-chip memory and data access scheduling strategies to address this issue, they still inevitably access features at irregular addresses from DRAM.

In this work, we propose LiGNN, a hardware-based solution that improves data locality by applying dropout and merge techniques during neighbor aggregation to accelerate GNN training.
Unlike conventional algorithm-level dropout methods that primarily aim to improve accuracy while overlooking hardware costs, LiGNN introduces a locality-aware feature dropout mechanism.  
This approach selectively drops node features with data locality awareness, effectively reducing irregular DRAM accesses without compromising model accuracy.
Moreover, by leveraging detailed knowledge of memory layout and organization—including critical alignment constraints—LiGNN strategically merges memory accesses during neighbor aggregation at the DRAM row level, guided by GNN-level semantics. This optimization significantly improves data locality with minimal additional cost.
Under the commonly adopted 0.5 dropout rate, LiGNN outperforms state-of-the-art methods, delivering a 1.48$\sim$3.02$\times$ speedup, reducing DRAM accesses by 34\%$\sim$55\%, and lowering DRAM row activations by 59\%$\sim$82\%, all while maintaining model accuracy.

\end{abstract}

%% file: tex/intro.tex
\section{Introduction}
Graph Neural Networks (GNNs) are a class of neural networks designed to operate on graph-structured data, effectively capturing relationships between entities~\cite{gnn_survey}. 
GNNs excel at tasks where data is naturally represented as nodes and edges, making them powerful for modeling complex systems.
GNNs have gained wide application in many critical fields such as electronic design automation (EDA)~\cite{GCN_EDA}, knowledge inference~\cite{gcn_knowledge_graph}, recommendation system~\cite{gcn_recommender_systems}, visual reasoning~\cite{gcn_visual_reasoning}, traffic prediction~\cite{google_map_gnn}, and so on~\cite{gnn_survey,survey_haiyang}.

A typical GNN model consists of multiple layers, each of which contains two main phases: aggregation and combination. 
In the aggregation phase, each node collects messages from its neighbors to generate an intermediate result, using an order-independent operator like element-wise sum or mean. This iteration process follows the sparse and irregular graph structure, where massive memory accesses with poor locality lead to significant performance degradation.
While in the combination phase, the intermediate result and features of the node are passed to a feedforward or dense layer, which shows regular execution pattern and high data reusability.
In all, the aggregation phase is memory-bound and time-consuming, thus becomes the main bottleneck of GNN models~\cite{understand_GCN,understand_HGNN,understand_hgnn_training, understand_gnn_survey}.

Memory layout and organization has been underestimated in optimization guided by application level information, especially when good alignments are provided. Without such knowledge, accessing neighbor feature in the aggregation phase is unaware of the underlying DRAM activities, thus missing the chance to optimize memory locality. Moreover, good alignments ensure simple mapping of neighbor to location in DRAM, making it practical to indirectly effect memory trace and performance using application level information.



Robustness, defined as the tolerance to data corruption or perturbation, is an important and widely observed property of GNNs. This robustness is often attributed to GNNs' intrinsic ability to aggregate information across topological structures, enabling them to mitigate the impact of localized noise or missing features.
To exploit this robustness, a series of dropout-based techniques have been proposed and applied specifically to various stages of the GNN aggregation process. For instance, methods such as DropOut \cite{DropOut}, DropEdge \cite{DropEdge}, DropNode \cite{DropNode}, and DropMessage \cite{DropMessage} strategically introduce randomness by selectively omitting nodes, edges, or message components during training. Empirical results across diverse datasets consistently demonstrate that these approaches improve model accuracy.


Therefore, not all accesses to node features are essential, presenting an opportunity to reorder and selectively drop certain irregular accesses to improve data locality and achieve speedup. However, algorithmic dropout methods inherently lack awareness of the DRAM working mechanism, making them incapable of making DRAM-friendly dropout decisions.

Accordingly, we propose LiGNN, a data locality extraction and enhancement hardware solution customized for dropout and merge in GNN training. 
LiGNN seats between DRAM and GNN training accelerator, intervening DRAM read accesses between them, with the help of application level information like aggregation edge lists.
We implement hardware dropout and merge by grouping and conditionally dropping DRAM read access to node features in the aggregation phase, considering locality. 
We primarily focus on DRAM, because GNNs have ultra low locality in the aggregation phase and almost all read requests are ultimately served by DRAM.
It is worth noting that our work is not a straight forward hardware implementation for previous algorithmic dropout efforts, but a novel design that enhance data locality inside DRAM following GNN robust nature.
To the best of our knowledge, no existing work has implemented dropout for GNNs as hardware-based filter seizing the opportunity provided by GNN robustness.

The key contributions of our work are as follows:
\begin{itemize}
\item We identify an unexplored approach to enhance data locality in GNN training and gain speedup by strategically dropping and merging DRAM access, in accordance with the robust nature of GNNs.

\item We introduce LiGNN, a hardware-based solution designed to enhance locality for DRAM accesses, specifically optimized to accelerate neighbor aggregation in GNN training.

\item We propose a locality-aware dropout mechanism with customized criteria for DRAM row filtering, aiming to balance dropout efficiency and the utilization of open DRAM rows, thereby streamlining row activations.
In addition, we introduce a locality-aware merging mechanism that clusters DRAM accesses based on both application-level and hardware-level information to further improve data locality.


\item We implement LiGNN in both RTL and cycle-accurate simulator, based on state-of-the-art (SOTA) GNN training accelerator GCNTrain\cite{GCNTrain} and evaluate it on several models, datasets and DRAM standards.
The results show that under classic 0.5 droprate, LiGNN achieves 1.48$\sim$3.02$\times$ speedup, reduces DRAM accesses by 34$\sim$55\% and DRAM row activation by 59$\sim$82\%, all without losing accuracy.

\end{itemize}

%% file: tex/background.tex
\section{Background}


\subsection{Graph Neural Network}
GNN is a kind of artificial neural networks that can process data representing as graphs. 
Using notations in Table \ref{tab:notation}, one layer of GNN can be formulated as follows:
\begin{table}[!t]
\caption{Notations used in this paper.} \label{tab:notation}
\centering
\renewcommand\arraystretch{1.2}

\begin{tabular}{c|c}
\hline
Notation & Explanation \\
\hline
$G=(V,E)$ & graph $G$ with vertices $V$ and edges $E$ \\
$v$ or $v_i$ & vertex $v$ (with index $i$) \\
$(i,j)$ or $e_{i,j}^k$ & ($k$-th feature of) edge from vertex $i$ to $j$ \\
$d_v, d_v^+, d_v^-$ & common, out, in degree of vertex $v$ \\
$N_v, N_v^+, N_v^-$ & common, out, in neighbor set of vertex $v$ \\
$x_v^k$ & input feature of vertex $v$ at $k$-th layer \\
$h_v^k$ & intermediate result of vertex $v$ at $k$-th layer \\
\hline
\end{tabular}
\end{table}
\[
    h_u^{k+1} = \phi\left(x_u^k, \displaystyle \bigoplus_{v \in N_u^-} \psi(x_u^k, x_v^k, e_{v,u}^k)\right),
\]
where $\psi$ and $\phi$ are message and update functions, respectively, and $\oplus$ is the aggregation operator.
The aggregation phase roughly corresponds to $\oplus$ and $\psi$, while combination phase is the rest $\phi$.
For example, the Graph Convolutional Network (GCN) \cite{GCN_paper} employs neighbor feature with coefficient as $\psi$, sum as $\oplus$, and matrix multiplication as $\phi$.
The Graph Attention Network (GAT) \cite{GAT_paper} uses a self-attention coefficient in $\psi$, and a multi-head attention in $\oplus$ and $\phi$.

Real-world graphs are extremely sparse and irregular. 
We define neighbor access irregularity for a traversal path as the mean of vertex index difference through it. 
Table \ref{tab:irregularity} demonstrates basic metrics, sparsity $\eta$ and irregularity $\xi$ of a sequential traversal path with two mean types (arithmetic and geometric) on several datasets.
It is clear that their sparsity are ultra high ($\eta>0.9999$) and their irregularity $\xi$ are just an order of magnitude lower than their number of vertices $|V|$.

\begin{table}[!htbp]
\caption{Graph irregularity.} \label{tab:irregularity}
    \centering
    \renewcommand\arraystretch{1.2}
    \begin{tabular}{c|ccc|cc}
    \hline
    Graph & $|V|$ & $|E|$ & $1-\eta$ & $\xi_A$ & $\xi_G$ \\
    \hline
    LiveJournal (LJ) & 4.8e6 & 6.9e7 & 2.9e-6 & 7.9e5 & 4.7e5 \\
    Orkut (OR) & 3.1e6 & 1.2e8 & 1.2e-5 & 8.1e5 & 5.8e5 \\
    Papers100M (PA) & 1.1e8 & 1.6e9 & 1.3e-7 & 3.2e7 & 2.2e7 \\
    \hline
    \end{tabular}
\end{table}

\subsection{DRAM Hierarchy and Working Mechanism}


In modern processor architecture, a memory controller communicates with memory modules over channels. A channel logically contains multiple memory modules (like DIMM), which in further consists of multiple DRAM chips. Following the hierarchy, rank, bank group, bank, row and finally column are introduced. The index of each hierarchy is directly specified by bits in given address, which the memory controller will provide on any read or write access. A row buffer acts as a placeholder when any row is desired, where the entire row is loaded and accessed in a smallest granularity called burst. Row buffer is exclusive to currently accessed row, which means accessing another row requires write-back and re-activation, which are slow and expensive.

Nowadays, modern architectures targeting neural networks mostly seek for efficient parallel execution. Accordingly, the DRAM bandwidth needs are urgent, hence they tend to employ many channels and maximize effective bandwidth of all channels by setting small interleaving and applying proper alignment. Such setup would scatter continuous address range to different rows, while keeping certain locality, which provides plenty room for our locality-aware hardware dropout and merge. 

%% file: tex/motivation.tex
\section{Motivation}

\subsection{Opportunity from GNN Robust Nature}
GNNs are robust enough to tolerant certain loss of input information. Algorithmic dropout \cite{DropOut, DropNode, DropEdge, DropMessage} is a popular robustness application, which enhances model accuracy by masking some input data. Those algorithmic dropout efforts reach higher accuracy and better applicability, suggesting that not all memory accesses are indispensable, as accuracy remains unaffected.
This insight highlights an opportunity for dropping irregular DRAM accesses in a locality-aware manner, to gain speedup without compromising model performance.

However, current GNN accelerators fail to take profit from robustness.
The aggregation phase of GNNs is characterized by irregular memory access patterns~\cite{understand_GCN,understand_HGNN}, posing significant challenges to execution platforms.
Previous GNN accelerators~\cite{HyGCN,AWB-GCN,FlowGNN,GCNTrain,HiHGNN} all assumed that every memory access during aggregation is essential, and have developed architectural optimizations based on this premise.

\subsection{Inefficiency of Algorithmic Dropout from Burst View}
\begin{figure}[!htbp]
    \centering
    \includegraphics[width=\linewidth]{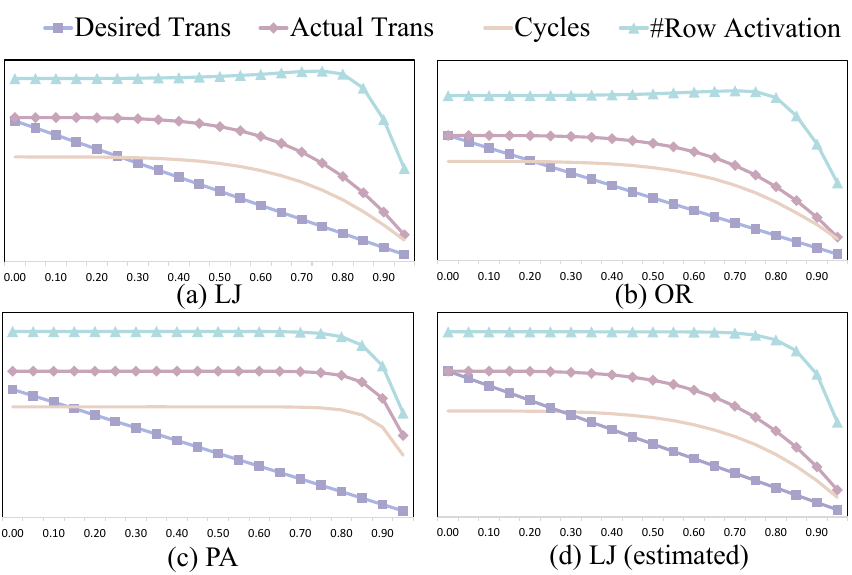}
    \caption{Effect of Algorithmic Dropout on DRAM Access Metrics with Different Drop Rates (LRU cache with 4K item, naive traversal, HBM).}
    \label{fig:prev_drop_perf}
\end{figure}

Existing GNN robustness application mainly focus on accuracy enhancement, hence we first characterize if such approach will efficiently boost performance.
All desired DRAM accesses finally resort to certain number of burst accesses, which is the minimal transaction in DRAM. Accordingly, we check if burst accesses are efficiently eliminated in aggregation with algorithmic dropout.
We consider DRAM accesses in the aggregation phase, in a naive traversal path. We focus on the initial aggregation and ignore other phases and layers, on an architecture with one level LRU cache (hosts 4K features) and HBM. 
The normalized execution cycles, desired/actual DRAM access amount and total row activation amount for three datasets are shown in Figure \ref{fig:prev_drop_perf}(a)(b)(c), where the horizontal axis is drop rate from 0 to 1. We define data amount really used by aggregation (pass algorithmic dropout) as desired amount, and the triggered burst transaction amount as actual amount.

The results shown that under algorithmic dropouts from previous studies, the desired amount decreases almost linearly as the drop rate increases, while the actual amount decreases much slowly. 
The total row activation amount is almost constant and finally reduces when drop rate $\alpha>0.8$. In all, desired amount reflects algorithmic dropout linearly, but actual amount and row activation amount are not efficiently eliminated.

In all, we conclude that, current application of GNN robust nature, namely algorithmic dropout, cannot efficiently eliminate DRAM access and boost performance, due to their primary accuracy-improving goal and lack of hardware knowledge. A hardware-based solution that utilizes GNN robust nature to reach performance enhancement is necessary and promising.

\subsection{Necessity and Benefit for Locality-aware Dropout and Merge}
\begin{figure}[!t]
    \centering
    \includegraphics[width=\linewidth]{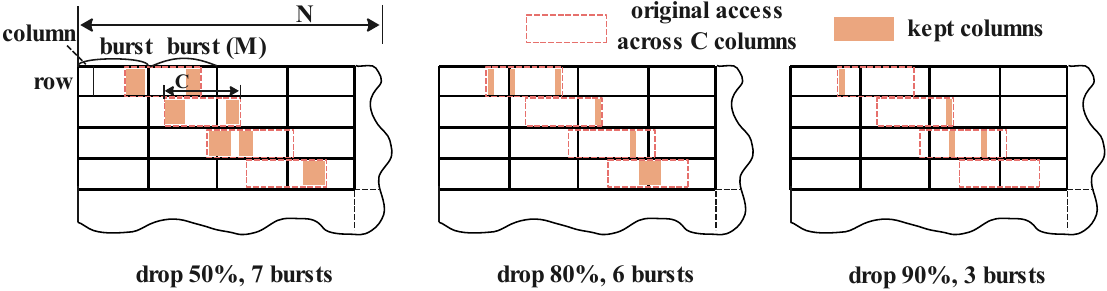}
    \caption{Typical DRAM Structure and Dropout Example.}
    \label{fig:bank_dropout}
\end{figure}

To estimate the theoretical speedup of locality-ware dropout over algorithmic dropout, we model how algorithmic dropout reflects on critical DRAM metrics like burst access or row activation amount. 
As shown in Figure \ref{fig:bank_dropout}, we assume a DRAM standard have $N$ columns per row, $M$ columns per burst, and $K$ elements per burst, where $M$ divides $N$. 
With interleaving, the input read accesses are $Q$ random ones each covering $C$ continuous columns in some row (typically $M<C<N$). 
We also assume the algorithmic dropout complies with Bernoulli$(\alpha)$ and ignore cache.
Ideally, the desired DRAM access amount is $QC(1-\alpha)$. 
However, the possibility of whole burst drop is $\alpha^{K}$, and the actual access amount will be $QC(1-\alpha^{K})$.
The inequality of drop rate to remaining bursts in Figure \ref{fig:bank_dropout} shows the burst-minimal DRAM characteristic.
Furthermore, the equivalent probability of a row skip is at most $\alpha^{CK/M}$, which can be ultra low.
Based on the modeling to algorithmic dropout, we draw estimated metrics in Figure \ref{fig:prev_drop_perf}(d), which fits real values and demonstrates the correctness of the model.

On the other hand, if effective locality-aware dropout is proposed so that the actual access amount is direct proportion of kept rate (namely $1-\alpha$), its theoretical performance can be estimated. As for actual access amount, we expect algorithmic dropout to be $(1-\alpha^K)/(1-\alpha)=1+\alpha+\cdots+\alpha^{K-1}$ times of locality-aware dropout. And for row activation amount, we may extend to $(1-\alpha^{CK/M})/(1-\alpha)$ times or even more, depending on different DRAM standards and interleaving.

\begin{figure}[!t]
    \centering
    \includegraphics[width=0.8\linewidth]{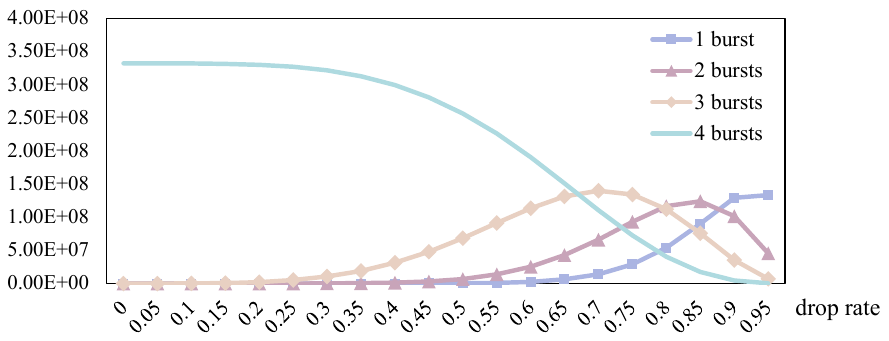}
    \caption{Distribution of Burst Access Amount per Row Open Session.}
    \label{fig:access_amount_per_session}
\end{figure}

With proper knowledge of DRAM organization and mapping, locality-aware dropout is totally practical, where burst access or row activation can be criteria of decision.
Thus, we see an unexplored way to introduce burst or even row as hardware dropout granularity, to save the execution platforms from heavy and exhausting DRAM accesses. 
Row granularity dropout is aggressive compared to burst. Its gain is obvious, due to the reduce of expensive row activation, so does its impact, since a row can easily goes over 8KB on most DRAM standards, which can leads to too coarse granularity and even accuracy loss. 
However, thanks to the maximizing parallelism setup of modern neural network accelerators, the effective bursts per row open session is much lower than the row size, avoiding too coarse granularity in a single row.
As shown in Figure \ref{fig:access_amount_per_session}, on Live Journal and GCN model with HBM, with alignment, the amount of actual access (burst) per row open session (max 4) is much less than number of bursts hosts in a row (64), which is normal due to graph irregularity and parallelism needs.

%% file: tex/design.tex
\section{Design of LiGNN}

Guided by the above observations, we propose LiGNN, a data locality extraction and enhancement solution, which offers DRAM burst and row as brand-new dropout and merge granularity in GNNs.

\begin{figure}[htb]
    \centering
    \includegraphics[page=1,width=\linewidth]{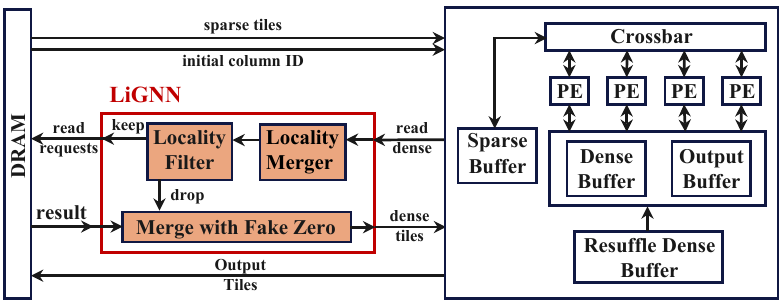}
    \caption{Architecture Overview of LiGNN based on GCNTrain.}
    \label{fig:arch_overview}
\end{figure}

Figure \ref{fig:arch_overview} provides an illustration of the proposed architecture, where inside the red rectangle is our LiGNN.
LiGNN does not depend on specific GNN training architectures, here we choose the SOTA GCNTrain \cite{GCNTrain} for example.
The GCNTrain-v3 abstracts SpMM operation and implement it with separate datapath for sparse and dense matrices, where the sparse tiles belong to graph structure and dense tiles are from feature and weight matrices. 
LiGNN acts as an agent for dense requests and tiles back, without touching sparse ones.
LiGNN will accept irregular read requests for dense matrix, conduct locality based filtering, send kept requests to and collect results from off-chip memory as usual, and finally output real accessed data and fake zero data to buffer of GCNTrain-v3 as dense tiles. 
In LiGNN, algorithmic dropout is substituted by hardware locality filter, while merge is achieved by hardware locality merger, which are both transparent for software.

\begin{algorithm}
\caption{locality\_aware\_group (burst dropout)}\label{alg:locality_order}
\begin{algorithmic}
\State \textbf{Input: } read request stream $S$, burst filter $B$, trigger $F$
\State \textbf{Output: } request table $T$ with address vector index
\While{$S$ is not empty}
\State $r \gets S$.get\_head\_and\_pop
\State $a \gets r$.address\_range
\For {\textbf{each} burst $b$ in $a$} 
    \If {$B$ drop $b$}
        \State \textbf{continue}
    \EndIf
    \State $v \gets b$.address\_vector
    \State $T[v] \gets T[v] \cup \{b\}$
    \State notify $F$ with $T$.size, $T[v]$.size, etc.
    \If {$F$.fire}
        \State \textbf{call} locality\_ordering\_output
    \EndIf
\EndFor
\EndWhile
\end{algorithmic}
\end{algorithm}

\subsection{Locality-aware Dropout}

\subsubsection{Locality-aware Ordering with Burst Dropout}
\begin{figure}[htb]
    \centering
    \includegraphics[page=2,width=0.9\linewidth]{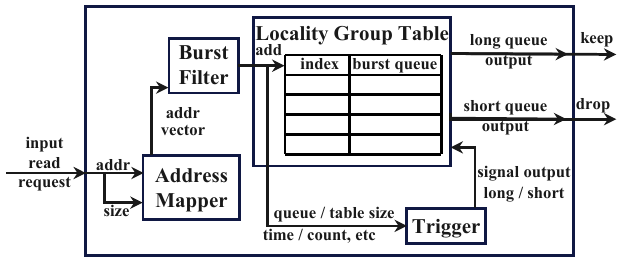}
    \caption{Architecture of Locality Filter.}
    \label{fig:arch_filter}
\end{figure}

To handle massive DRAM read requests, as shown in Alogrithm \ref{alg:locality_order} and Figure \ref{fig:arch_filter}, we propose a locality-aware grouping and ordering policy to categorize and compare them efficiently. We use an address-vector-indexed table to temporarily store the grouped accesses for further decision-making, allowing for selection at the granularity of bursts or rows. 
When processing an incoming read request for a node feature, we first retrieve its address range from model information and generate the corresponding actual accesses (bursts) to that range, taking into account DRAM organization and mapping, with address translation applied if necessary. For each burst, we apply a filter $B$ to determine whether it should be dropped, considering factors such as its effective ratio (since part of the burst may be masked by dropout) or load balancing.

After passing through the burst filter, burst accesses are grouped into the locality group table (LGT) based on their address vector, following a DRAM hierarchy. The LGT is implemented as a content-addressable memory (CAM), where the key is the row identifier and the value is a FIFO queue for bursts. During the grouping process, a predefined custom trigger $F$ is notified with relevant information such as the size of the LGT (or its items), elapsed time, or compute engine utilization. Based on this information, the trigger can decide when to output data. If no trigger is defined, all bursts that pass filter $B$ are retained, while those that do not pass are dropped, rendering the LGT unnecessary in such cases.

\begin{algorithm}[!t]
\caption{locality\_ordering\_output (row dropout)}\label{alg:output_size}
\begin{algorithmic}
\State \textbf{Input: } request table $T$ with address vector index
\State \textbf{Input: } desired size $n$, dropping rate $\alpha\in(0,1)$, criteria $C$
\State \textbf{Output: } kept request queue $K$, dropped request queue $D$
\State \textbf{Init: } $\delta \gets 0$
\State $k \gets 0, d\gets 0$
\While{$T$ is not empty \textbf{and} $k + d < n$}
\If{$\delta + (k+d)\alpha - d > 0$}
    \State move shortest queue $x$ in $T$ to $D$
    \State $d \gets d + x$.size
\Else
    \State move the longest queue $x$ fits $C$ in $T$ to $K$
    \State $k \gets k + x$.size
\EndIf
\EndWhile
\State $\delta \gets \delta + (k+d)\alpha - d$
\end{algorithmic}
\end{algorithm}

\subsubsection{DRAM Row Integrity Policy for Row Dropout}
To reach row granularity dropout with custom trigger, a configurable drop rate $\alpha\in(0,1)$ is required and a persist balance $\delta$ is maintained, whose sign bit determines whether next step is to drop or to keep. 
We also maintain a kept and dropped number in every call, for count and stop condition. 
If current state is to-drop, we drop the shortest queue in LGT, which is row granularity. 
Otherwise, we find the longest queue that fits custom criteria $C$ and keep it. 
A random one is picked if multiple shortest/longest queues exist.
Criteria $C$ is set for needs like channel balancing or row-policy preference. For example, we can even cancel the queue size requirement and treat all queues equally.
The process is ended when total output amount is reached and afterwards, the persist $\delta$ is updated. 
The comparison process of queue size is implemented in comparison complete binary tree style, where the values and indices are compared by trees to find large and small one (random if equal). 
Using the final output indices, we perform output operation over the corresponding FIFOs (queue in LGT).

\subsection{Locality-aware Merging}

\begin{figure}[t]
    \centering
    \includegraphics[page=3,width=0.9\linewidth]{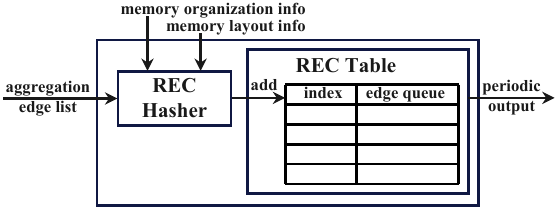}
    \caption{Architecture of Locality Merger.}
    \label{fig:arch_merger}
\end{figure}

With locality-aware dropout, partial irregular DRAM accesses are eliminated and overall locality is improved.
To further utilize the application level information of GNN workloads, we propose a locality-aware merging method to rearrange the read requests in aggregation phase, as shown in Figure \ref{fig:arch_merger}.
This method takes the edge list and necessary memory organization and layout info as configurable input, 
forms a row equivalence class (REC) hasher, where aggregation neighbors are essentially \textit{hashed} with their ID to indicate their feature data location in DRAM rows.
In other words, two neighbors share DRAM rows if their row-hash value equal.
At runtime, the REC hasher categorize edges from aggregation edge list with their row-hash value and put them in REC table, where edge queues are periodically output.
With proper alignment and DRAM organization, the hashing method simplifies to rearrangement by bit operation of vertex indices.
Different from the locality-aware dropout method, the proposed merging method essentially reorders the read requests, with help of application information like edge list, while keeping all requests intact.

To illustrate the core process of hash under proper alignment and DRAM organization, we assume proper alignment (power of 2) of feature matrices and feature vectors with the knowledge of memory physical / logical layout. 
Based on the layouts and alignment, the start address of any neighbor feature can be calculated. 
Also, two neighbor features share any DRAM rows if and only if their start address share one DRAM row, which simplifies REC hasher.
For example, with data type of \texttt{float32 (4B)}, feature matrix start address \texttt{S} aligned to \texttt{M KB (M=4,8,16)} and feature vector length \texttt{N (N=64,128,256,512)}, the start address of feature of vertex $v$ will be \texttt{S+N*4*v=k*M*1024+N*4*v}, with \texttt{k} being some integer. 
Based on the start address and memory physical layout, the DRAM row start address can be deduced. 
Take a typical HBM organization as an example, where the column and transmit (burst) bits in address vector are \texttt{13:9} and \texttt{5:0} respectively, while the parameters being \texttt{M=4, N=256}. 
Then the DRAM row start address of feature of vertex $v$ will be \texttt{16384*(k>>2+v>>3)}, hence vertices $v,u$ share any DRAM rows will be equivalent to \texttt{v\&\char`~7 == u\&\char`~7}, where merge can happen if the condition meets. 
Via this simplification from alignments, locality-aware merging is fast and cheap to implement.

\subsection{Integration Discussion}

How LiGNN integrates with other components is essential for practical usage.
The memory layouts can be passed on configuration and will not change. The proper memory alignments can also be easily assured by software/hardware configurations.
As for interface, besides adding dedicated signals, to utilize the QoS field of AXI protocol and a special QoS value as ``droppable" is quite practical.
If a bit droppable flag for every DRAM access is not feasible, the pre-configured address ranges are another way to distinguish between normal and droppable accesses.
Sometimes the dropout mask is needed for other steps like backward pass, hence every address range or read request may also carry address info of its mask for LiGNN to write, otherwise sending flags back aside read results is also viable. 
The dropout mask is usually single-bit boolean and stored continuously, like an edge feature. Hence write to the mask will show good locality, in contrast to reading the feature data.
The dropout scaling step, namely multiply by constant $1/(1-\alpha)$, is not done by LiGNN but done by compute unit, which can be easily inserted somewhere in the whole process.

\begin{table}[!htbp]
\renewcommand\arraystretch{1.2}
\setlength\tabcolsep{4pt}%
    \centering
    \caption{Detailed Parameters for \texttt{LG-\{A,B,R,S,T\}}.}
    \label{tab:parameter_of_LG-var}
    \begin{tabular}{c|ccccc}
    \hline
    Name & Trigger Fire & Burst Filter & Row Filter & LGT size & Merge \\
    \hline
    \texttt{LG-A} & N.A. & Element-wise & N.A. & N.A. & N.A. \\
    \texttt{LG-B} & N.A. & Yes & N.A. & N.A. & No \\
    \texttt{LG-R} & Feature & Optional & Yes & 16x16 & No \\
    \texttt{LG-S} & Custom & Optional & Yes & 64x32 & No \\
    \texttt{LG-T} & Custom & Optional & Yes & 64x32 & Yes \\
    \hline
    \end{tabular}
\end{table}

We implement several variants of LiGNN and name them by suffices \texttt{A}, \texttt{B}, \texttt{R}, \texttt{S}, \texttt{T}, while the prefix is \texttt{LG} for abbreviation of LiGNN. 
We first define \texttt{LG-A} as algorithmic dropout baseline, just for comparison.
We then define \texttt{LG-B} as burst filter only version, where actually the LGT and trigger $F$ do not exist.
We also define \texttt{LG-R} as row filter version, namely LGT with trigger $F$ and no burst filter, where the trigger fires on every feature read request. 
We expand the row filter scheduling range at decision (or trigger firing interval) to custom interval like certain number of features or by utilization of other units, and name it as \texttt{LG-S}. 
Finally, we enable locality-aware merging to \texttt{LG-S} and get \texttt{LG-T}.
In all, from \texttt{LG-B}, \texttt{LG-R}, \texttt{LG-S} to \texttt{LG-T}, the complexity and gain both increase.
The detailed parameters for \texttt{LG-\{A,B,R,S,T\}} is shown in Table \ref{tab:parameter_of_LG-var}. 

%% file: tex/evaluation.tex
\section{Evaluation}

\subsection{Methodologies}
\subsubsection{Evaluation Tools}
Based on SOTA GCN training accelerator GCNTrain~\cite{GCNTrain}, we implement a cycle-accurate simulator to measure the execution time in number of cycles. The critical component, DRAM, is simulated by Ramulator \cite{ramulator}, which supports cycle-accurate model and energy estimation for various DRAM standards. 
To measure critical paths in LiGNN, we implement the core component, namely LGT with input/output logic in RTL and synthesize it in Verilog. We use the Synopsys Design Compiler with the TSMC 12 \textit{nm} standard VT library for the synthesis and estimate the power consumption using Synopsys PrimeTime PX. The critical path, which is in CAM lookup, has a delay of 0.81 \textit{ns} including the setup and hold time, enabling smooth executing of LiGNN at 1 GHz clock frequency. The access latency, energy, and area of on-chip memory like CAM and FIFO are estimated using Synopsys DesignWare Memory Compiler.

\subsubsection{Baseline and System Configurations}
We choose \texttt{LG-A} as baseline and \texttt{LG-\{B,R,S,T\}} as gradually improved design.
The burst filters employ distribution in previous algorithmic dropout works. 
We choose HBM, DDR4 and GDDR5 as three representative DRAM standards, from various ones shown in Table \ref{tab:dram_specs}. We focus on execution cycles, desired/actual DRAM access amount, memory locality and total row activation amount.

\begin{table}[!t]
\centering
\renewcommand\arraystretch{1.2}
\setlength\tabcolsep{4pt}%
\caption{Typical Specifications of Common DRAM Standards.}
\label{tab:dram_specs}
\begin{tabular}{l|c|c|c|c|c}
\hline
Standard & \thead{Frequency\\(MHz)} & \thead{Bandwidth\\(GB/s)} & \thead{Columns\\Per Row} & \thead{Column\\Size (bits)} & \thead{Burst} \\
\hline
DDR3 & 400--1066 & Up to 17 & 1K & 64 & 8 \\
    DDR4 & 1600--3200 & Up to 25.6 & 1K & 64 & 8 \\
    GDDR5 & 1750--2000 & Up to 256 & 1K & 32 & 8/16 \\
    GDDR6 & 2500--3500 & Up to 768 & 1K & 32 & 16 \\
    LPDDR4 & 1600--4266 & Up to 34 & 1K & 64 & 16 \\
    LPDDR5 & 2750--6400 & Up to 51.2 & 1K & 64 & 16 \\
    HBM & 500 & 128 & 128 & 128 & 2 \\
    HBM2 & 1000--1200 & Up to 307 & 64 & 128 & 2 \\
\hline
\end{tabular}

\end{table}

\subsubsection{Workloads} We choose the famous GCN~\cite{GCN_paper}, GraphSAGE~\cite{GraphSage} and GIN~\cite{GIN} with two layers as three main models, since the base architecture GCNTrain is for GCN training. The dataset used is listed in Table \ref{tab:irregularity}. We select a dropout probability $\alpha$ from 0 to 1 with a step of 0.1. 

\subsection{Overall Results}
We evaluate the speedup, DRAM access amount, data locality, area and power of \texttt{LG-T} and \texttt{LG-A}, on three datasets with three GNN models and HBM standard.

\subsubsection{Speedup}
\begin{figure}[!htbp]
    \centering
    \includegraphics[page=1, width=\linewidth]{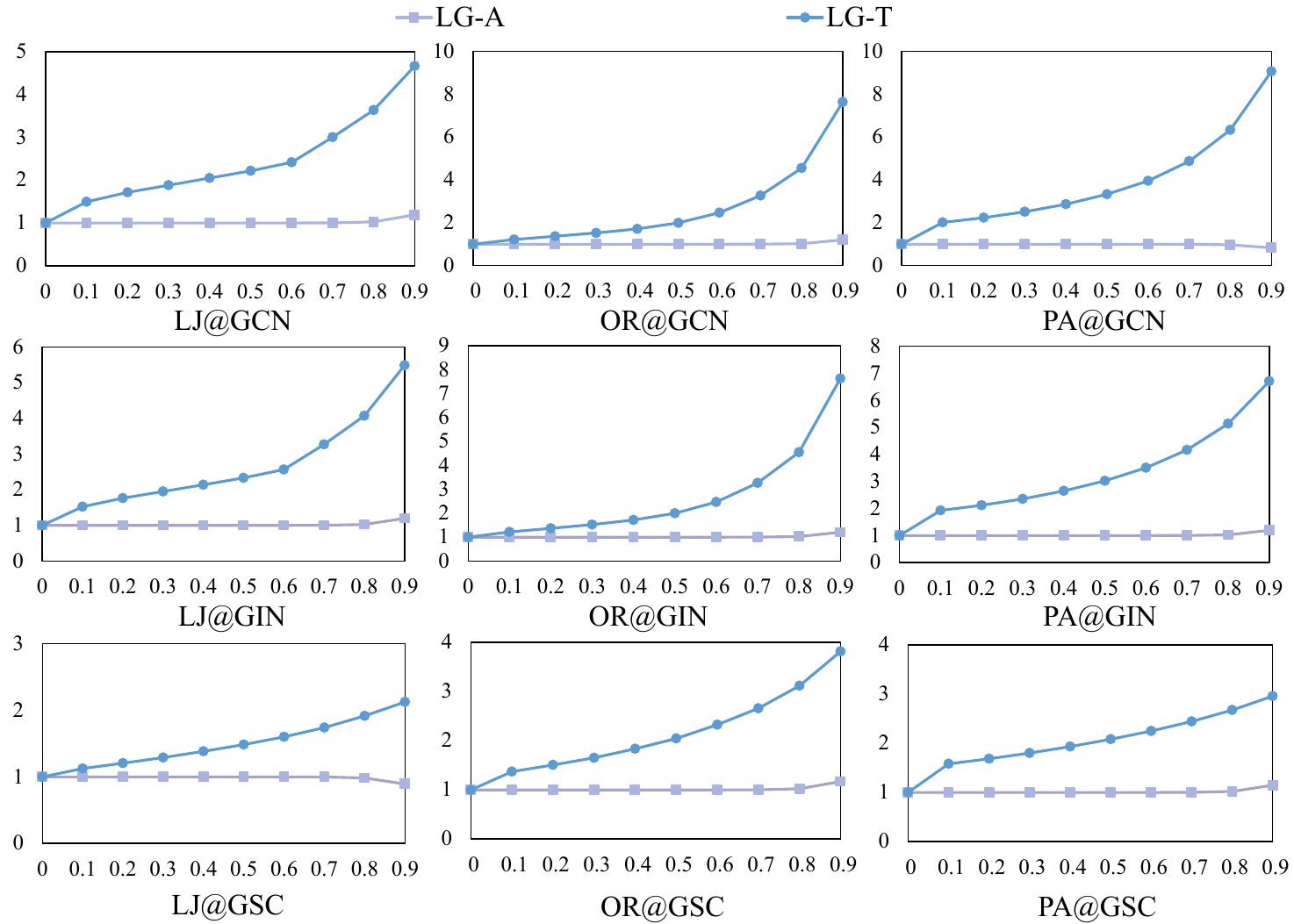}
    \caption{Speedup of \texttt{LG-T} and \texttt{LG-A} on different droprates.}
    \label{fig:overall_speedup}
\end{figure}

Figure \ref{fig:overall_speedup} compares the speedup of \texttt{LG-T} and \texttt{LG-A}, against non-dropout execution, across different $\alpha$.
We see that \texttt{LG-T} reaches good speedup than \texttt{LG-A} (specifically, 1.48$\sim$3.02$\times$ at $\alpha$=0.5), effectively seizing the opportunity of dropped and merged read accesses. Moreover, \texttt{LG-T} reaches near-linear performance gain as drop rate increasing, while \texttt{LG-A} gets few improvement.

\subsubsection{DRAM Access Amount}
\begin{figure}[!htbp]
    \centering
    \includegraphics[page=2, width=\linewidth]{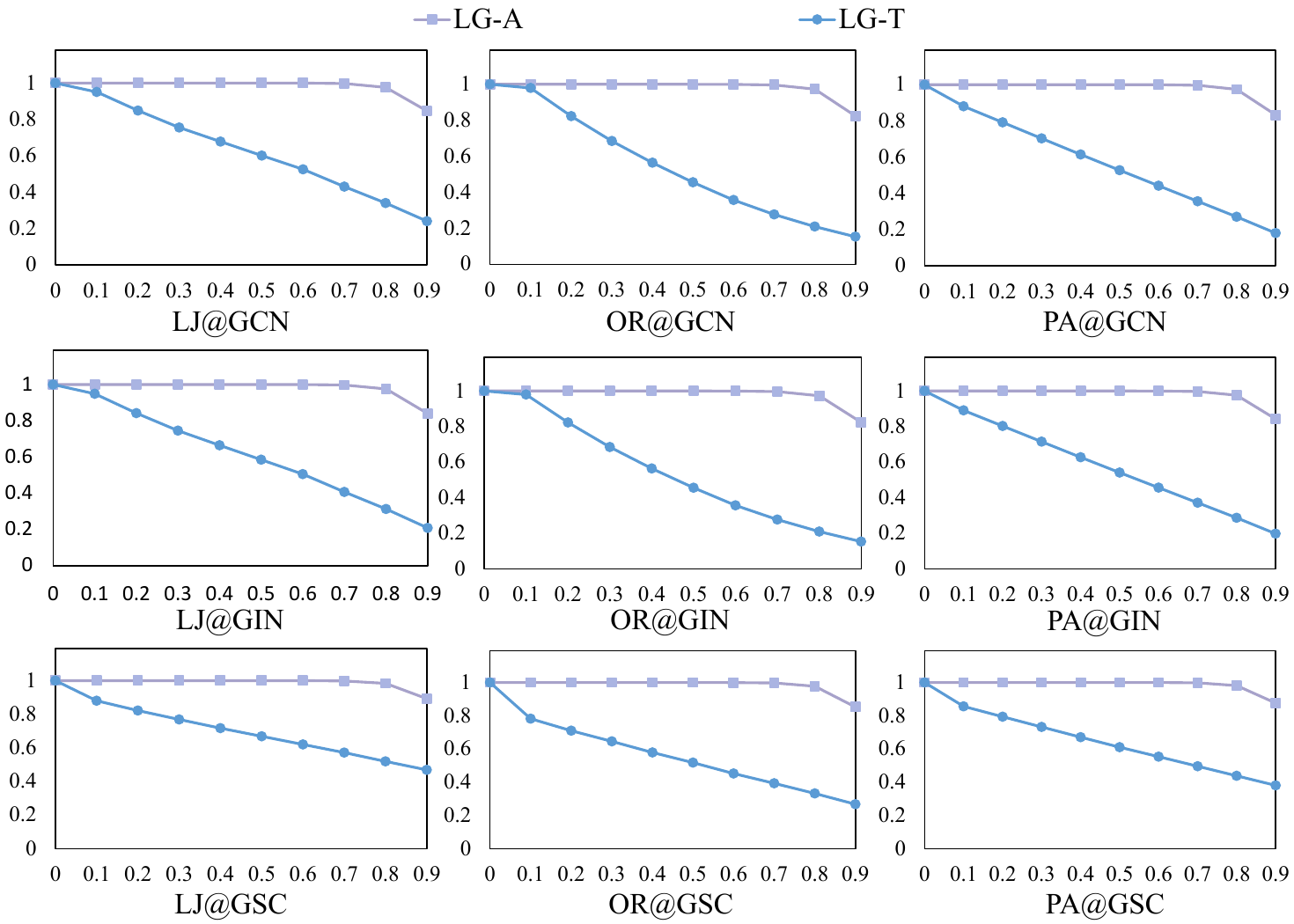}
    \caption{DRAM Access Amount of \texttt{LG-T} and \texttt{LG-A} on different droprates.}
    \label{fig:overall_amount}
\end{figure}

Figure \ref{fig:overall_amount} compares the DRAM access amount of \texttt{LG-T} and \texttt{LG-A}, against non-dropout execution, across different $\alpha$. 
We see that \texttt{LG-A} reduces little DRAM access, regardless of $\alpha$ value, while \texttt{LG-T} reaches approximately sub-linear reduction (specifically, 34\%$\sim$55\% reduction at $\alpha$=0.5).

\subsubsection{Data Locality} 
\begin{figure}[!htbp]
    \centering
    \includegraphics[page=3, width=\linewidth]{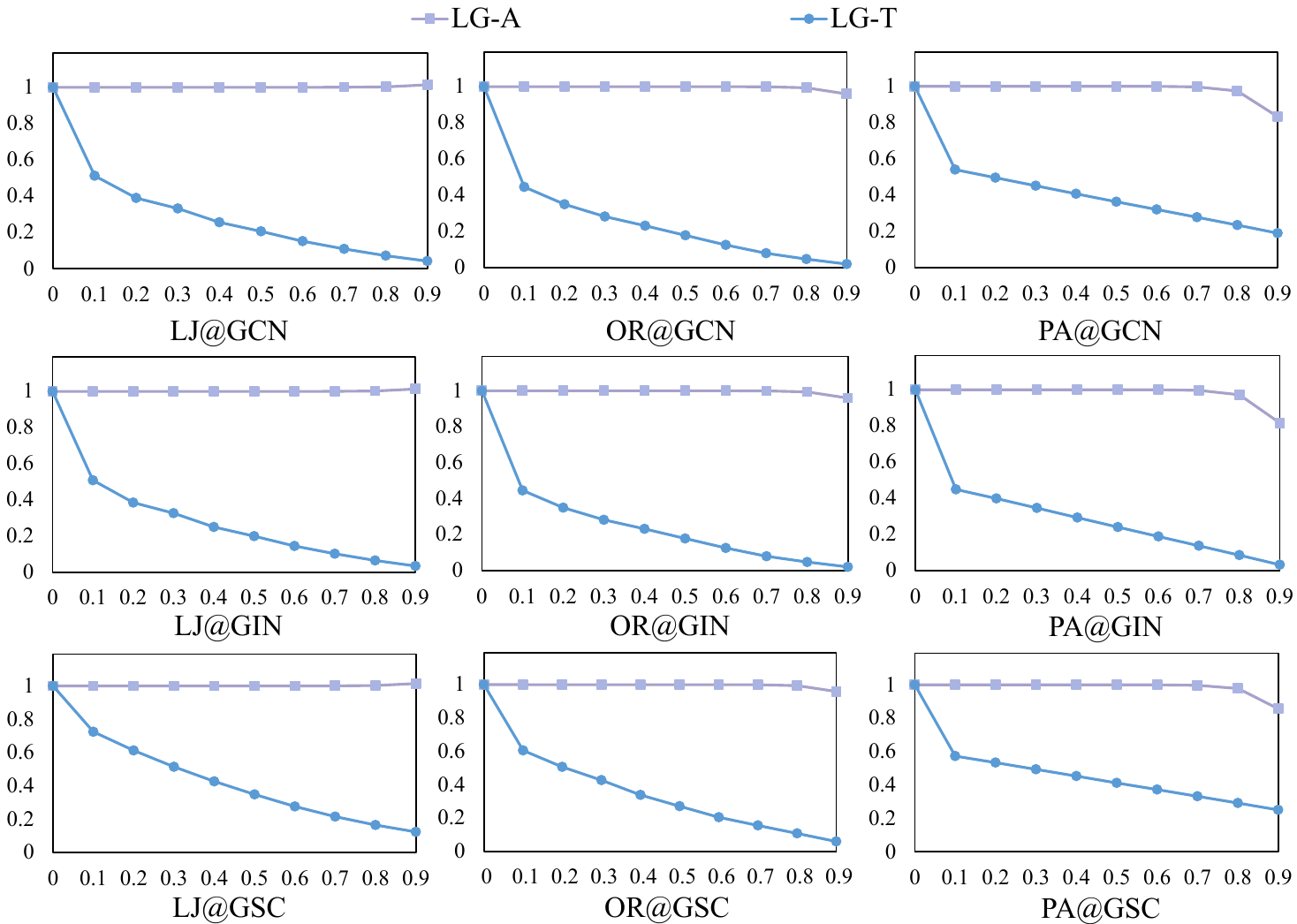}
    \caption{DRAM row activation amount of \texttt{LG-T} and \texttt{LG-A} on different droprates.}
    \label{fig:overall_locality}
\end{figure}

DRAM row activation reflects the data locality via its total amount and consumes palpable energy.
Figure \ref{fig:overall_locality} compares the DRAM row activation amount of \texttt{LG-T} and \texttt{LG-A}, against non-dropout execution, across different $\alpha$. 
With \texttt{LG-T}, considerable amount of row activation is eliminated, (specifically, 59\%$\sim$82\% reduction at $\alpha$=0.5) while \texttt{LG-A} has little effect.
Hence, combining locality-aware merging and dropout can effectively improve data locality.

\subsubsection{Area and Power}


Overall, LiGNN incurs a maximum area overhead of 0.04 mm$^2$ and power consumption of 21 mW under the TSMC 12 nm process.
The LGT is implemented as a CAM+FIFO structure, occupying approximately 0.006 mm$^2$ and 0.03 mm$^2$, and consuming up to 3 mW and 15 mW for \texttt{LG-{R,S}}, respectively.
The REC hasher, implemented as combinational logic based on integer bit-wise operations, has negligible area and power overhead compared to storage.
The REC table can also be implemented using a CAM+FIFO structure, with an area of approximately 0.01 mm$^2$ and a maximum power consumption of 6 mW.
Overall, compared to the original GCNTrain design—which occupies 0.9 mm$^2$ and consumes 143 mW under the TSMC 28 nm process—as well as other GNN accelerators, LiGNN incurs only a small fraction of the area and power, introducing minimal hardware overhead.

\subsection{Effect Analysis for Locality-aware Dropout}

\subsubsection{Speedup}
\begin{figure}[!htb]
    \centering
    \includegraphics[width=\linewidth]{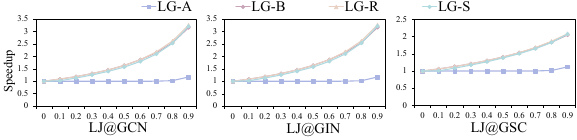}
    \caption{Speedup on Different Droprates.}
    \label{fig:eval_speedup_amount}
\end{figure}

Figure \ref{fig:eval_speedup_amount} compares the speedup of the proposed \texttt{LG-\{B,R,S\}} and baseline \texttt{LG-A} over non-dropout value.
We see that as $\alpha$ goes up, baseline \texttt{LG-A} reaches little speedup while \texttt{LG-\{B,R,S\}} achieve super-linear speedup.
Specifically, at $\alpha=0.5$, \texttt{LG-\{B,R,S\}} reaches 1.38$\sim$1.73$\times$
speedup, demonstrating the effect of locality-aware dropout.

\subsubsection{DRAM Access Amount} 
\begin{figure}[!htb]
    \centering
    \includegraphics[width=\linewidth]{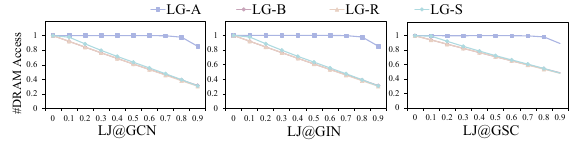}
    \caption{Normalized Actual DRAM Access Amount on Different Droprates.}
    \label{fig:eval_acc_amount}
\end{figure}

Figure \ref{fig:eval_acc_amount} compares the DRAM access amount over non-dropout value on LJ dataset. As $\alpha$ goes up, baseline \texttt{LG-A} reduces little DRAM access, while \texttt{LG-\{B,R,S\}} achieve linear reduction, showing the effect of DRAM burst granularity dropout.

\begin{figure}[!htb]
    \centering
    \includegraphics[width=\linewidth]{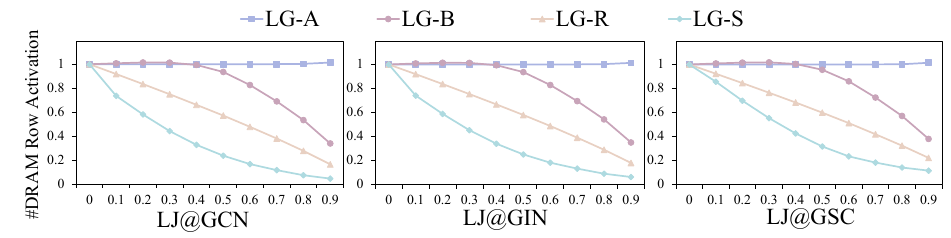}
    \caption{Normalized DRAM Row Activation Amount on Different Droprates.}
    \label{fig:eval_act_amount}
\end{figure}

\subsubsection{Data Locality}
DRAM row activation reflects the data locality via its total amount and consumes palpable energy.
Figure \ref{fig:eval_act_amount} compares the DRAM row activation amount over different droprates. We see that \texttt{LG-\{A,B,R,S\}} successively expose smaller row activation amount, which means better locality and lower energy consumption. Specifically, \texttt{LG-R} is nearest to a linear correlation of $\alpha$ while \texttt{LG-S} is a little super-linear, revealing the effect of DRAM row integrity policy.

\subsubsection{Exploration}
\begin{figure}[!htb]
    \centering
    \includegraphics[width=\linewidth]{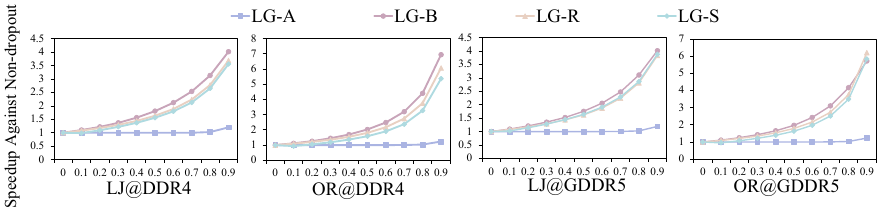}
    \caption{Speedup over DDR4 and GDDR5.}
    \label{fig:eval_mem_speedup}
\end{figure}

\begin{figure}[!htb]
    \centering
    \includegraphics[width=\linewidth]{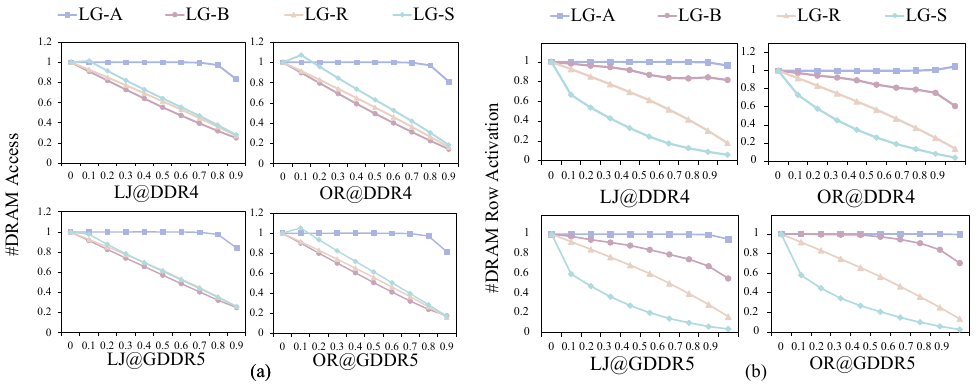}
    \caption{DRAM Access and Row Activation Amount over DDR4 and GDDR5.}
    \label{fig:eval_mem_acc_row}
\end{figure}

We evaluate LiGNN on DDR4 and GDDR5 with GCN model. Figure \ref{fig:eval_mem_speedup} and Figure \ref{fig:eval_mem_acc_row} compare speedup, DRAM access and row activation amount respectively, and are both similar to that on HBM, showing the good adaptability of LiGNN.

Numerous algorithmic dropout works have demonstrated that proper dropout will not degrade model accuracy, but help reach higher accuracy.
We also analyze the effect of burst or row dropout in LiGNN on model accuracy using PyG and DGL, with a two-layer GCN model. As shown in Table \ref{tab:dropout_accuracy}, burst or row dropout show no significant loss of model accuracy.

\begin{table}[!htb]
\renewcommand\arraystretch{1.2}
    \centering
    \caption{Effect of Burst and Row Dropout on Model Accuracy.}
    \label{tab:dropout_accuracy}
    \begin{tabular}{c|cccc}
        \hline
        Droprate & 0 & 0.1 & 0.2 & 0.5 \\
        \hline
        Burst Dropout & 0.77 & 0.758 & 0.764 & 0.757 \\
        Row Dropout & 0.77 & 0.76 & 0.768 & 0.762  \\
        \hline
    \end{tabular}
\end{table}

\subsection{Effect Analysis for Locality-aware Merging}

We evaluate locality-aware merging (LM) separately with non-merge (NM) version, on Live Journal with GCN model and HBM. 
The number of concurrent access (\texttt{Access}), on-chip memory capacity (\texttt{Capacity}, in number of node features), feature length (\texttt{Flen}) and schedule range (\texttt{Range}) are various. 
In non-merge version, least recently used (LRU) algorithm is used for on-chip memory replacement.

\subsubsection{Speedup}
\begin{figure}[!htb]
    \centering
    \includegraphics[width=1\linewidth]{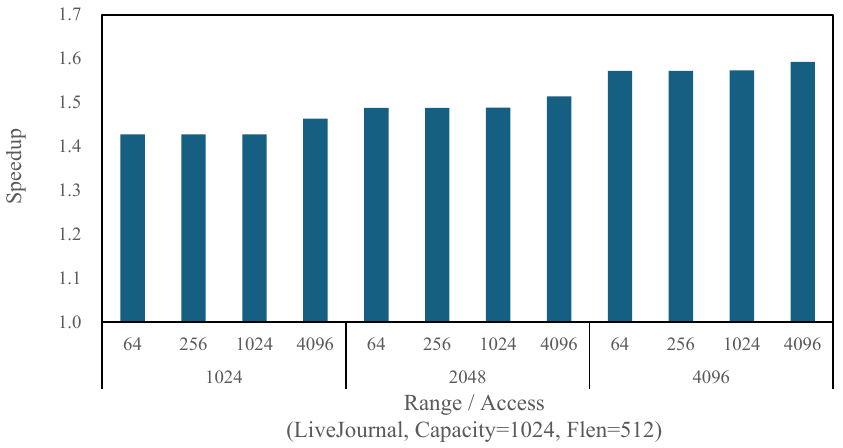}
    \caption{Speedup of LM over NM on Live Journal with Various \texttt{Range} and \texttt{Access}.}
    \label{fig:onchip_time_lj_2}
\end{figure}

Figure \ref{fig:onchip_time_lj_2} shows the speedup of locality-aware merging over non-merge version on Live Journal, with various \texttt{range} and \texttt{access}. 
The locality-aware merge version gains 1.43$\sim$1.59$\times$ speedup.

\subsubsection{Data Locality}

\begin{figure}[!htb]
    \centering
    \includegraphics[width=1\linewidth]{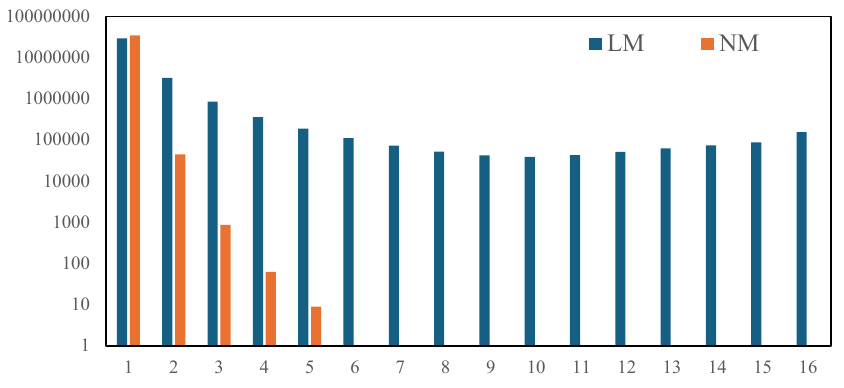}
    \caption{DRAM Row Session Size Distribution of LM on Live Journal.}
    \label{fig:lj-row-size}
\end{figure}

Figure \ref{fig:lj-row-size} shows the DRAM row session size distribution of locality-aware merging and non-merge version on Live Journal, with \texttt{Flen=512, Capacity=1024, Range=1024, Access=1024}. 
By DRAM row session size, we refer to the number of burst access in a open row session. Generally, larger row session size means better data locality.
We see that locality-aware merging effectively reduce count of row session with size one, and pushing them to bigger row sessions, essentially improving data locality.
It is also admitted that graph structures are so sparse and irregular that locality-aware merging can only merge a small part of memory accesses, hence row sessions of size one still dominates.

\subsubsection{DRAM Access Breakdown}

\begin{figure}[!htb]
    \centering
    \includegraphics[width=1\linewidth]{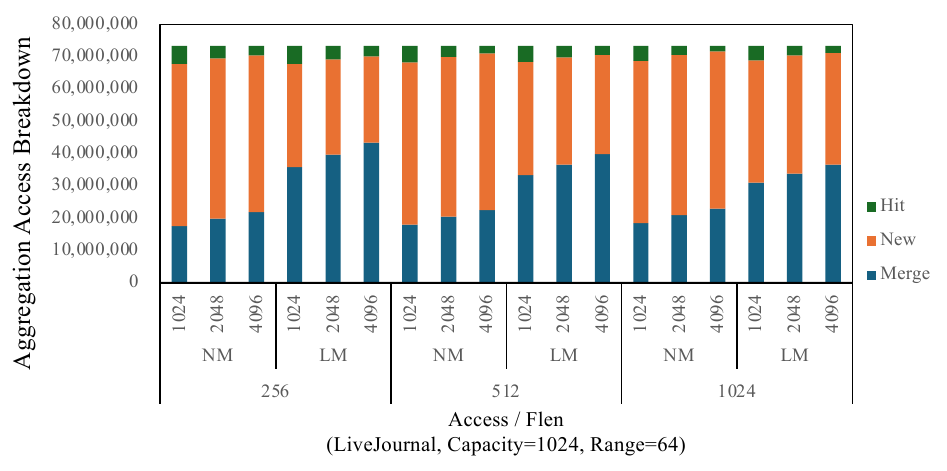}
    \caption{DRAM Access Breakdown on LiveJournal with various \texttt{Access} and \texttt{Flen}.}
    \label{fig:lj_breakdown}
\end{figure}

Figure \ref{fig:lj_breakdown} shows the relationship between DRAM access breakdown with \texttt{access} and \texttt{flen} on the Live Journal dataset, where \texttt{capacity} and \texttt{range} are fixed values.
The overall demand for reading and writing feature vectors is basically constant, which approximately equal to the sum of the vertex number and edge number of the graph dataset.
We classify all memory access as three types, namely hit, new and merge. By hit we mean accesses served directly by on-chip memory. While by new and merge, we refer to accesses that served by DRAM 
in a new row session, or in an existing session, respectively.
We see that locality-aware merging can effectively merge accesses.
Although overall DRAM access amounts are almost the same, the more merged, the better performance gained.

\subsubsection{Exploration}

\begin{figure}[!htb]
    \centering
    \includegraphics[width=1\linewidth]{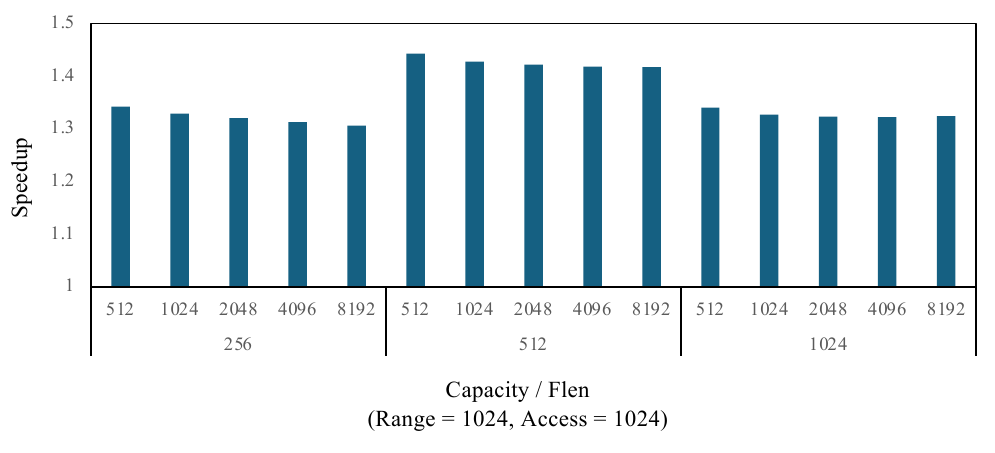}
    \caption{Speedup of LM over NM on LiveJournal with various \texttt{Capacity} and \texttt{Flen}.}
    \label{fig:lj_exp_1}
\end{figure}

We also conduct more exploration experiments on the Live Journal dataset. 
Figure \ref{fig:lj_exp_1} shows the speedup of locality-aware merging over non-merge version, with various \texttt{capacity} and \texttt{flen}. 
The locality-aware merge version gains 1.30$\sim$1.44$\times$ speedup. 
It is also shown that under \texttt{range=1024} and \texttt{access=1024}, the speedup is highest when \texttt{flen=512}. The reason is that too short or too long feature has different impacts on DRAM than that of proper length, hence proper but not always better configurations are needed.

\begin{figure}[!htb]
    \centering
    \includegraphics[width=1\linewidth]{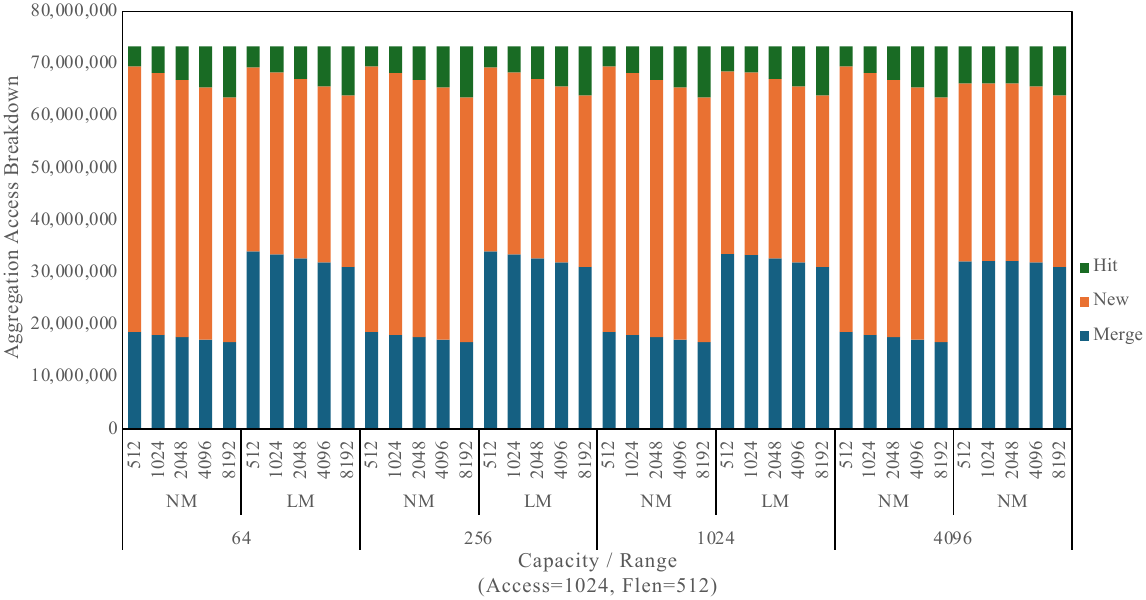}
    \caption{DRAM Access Breakdown on LiveJournal with various \texttt{Capacity} and \texttt{Range}.}
    \label{fig:lj_exp_2}
\end{figure}

Figure \ref{fig:lj_exp_2} shows the DRAM access breakdown of locality-aware merging over non-merge version, with various \texttt{capacity} and \texttt{range}.
It is shown that locality-aware merge version effectively merge more access that non-merge version, similar to that on \texttt{capacity=1024} and \texttt{range=64}, showing the good scalability of our design.

%% file: tex/related_work.tex
\section{Related Work}

\textbf{GNN Characterization.} A growing body of work has focused on the characterization and analysis of GNNs, which is critical for identifying performance bottlenecks and informing the design of efficient GNN accelerators~\cite{understand_GCN,understand_gnn_survey,understand_HGNN,understand_hgnn_training,understanding_distributed_gnn_training_gpu,GNN_architectural_implications_GPU_TPU,GNN_Mark_gnn_training_benchmark_gpus,empirical_analysis_gnn_runtime_gpus,framework_analysis_time_memory_gnn,dynamic_graph_inf_cpu_gpu,defense_gnn_gpu}.
For instance, the study~\cite{understand_GCN} provides a detailed bottleneck analysis of GCN inference on GPUs, highlighting key performance constraints. Similarly, GNNMark~\cite{GNN_Mark_gnn_training_benchmark_gpus} introduces a comprehensive benchmark suite designed to evaluate GNN training performance on GPU platforms, offering empirical insights into resource utilization, runtime behavior, and scalability.

\textbf{GNN Accelerators.} Existing GNN accelerators~\cite{HyGCN, AWB-GCN, GCNTrain, FlowGNN, HiHGNN, MetaNMP, GDR-HGNN, GraphACT, SiHGNN, mini_batch_GNN_inference_cpu_fpga, MultiGCN, rubik} reach great success in boosting performance of GNN execution and reduce energy consumption. However, they all operate under the assumption that every DRAM access during aggregation is essential, and have developed architectural optimizations based on this premise.
Recent GAT acceleration work~\cite{TopK_GAT,ADE-HGNN} skip computation and access for low-attention value vertices, but such approximate computing approach still originates from algorithm improvement but not hardware demands.

\textbf{GNN Algorthmic Dropout.}
GNNs are robust enough that they can tolerate proper loss of information, which 
has been utilized by various algorithmic efforts. Random dropping is firstly introduced in \cite{DropOut}, and proved effective by works \cite{lh_nn,drop_train_svm,dropout_overfit}. Work \cite{dropout_regul} proves that corrupted features are equivalent to L2-type regularization. Work \cite{dropout_ada_regul} shows deeper and more clear relationship between dropout regularizer and L2 regularizer.
As GNN comes up, random dropping has been generalized to graph data, by several representative works. DropOut \cite{DropOut}, DropNode \cite{DropNode}, DropEdge \cite{DropEdge}, DropMessage \cite{DropMessage} propose dropout in different granularity from element to whole feature and successfully improve the model accuracy.

However, random dropping methods are purely algorithmic solutions aimed primarily at improving accuracy rather than enhancing performance. While they may incidentally reduce desired DRAM access, their lack of memory-awareness limits their effectiveness in minimizing actual DRAM access. Therefore, a hardware-based dropout solution is essential for accelerating GNN training.

%% file: tex/conclusion.tex
\section{Conclusion}

This work identifies an opportunity for improving data locality and gain speedup, with reasonable use of GNN robust nature and DRAM timing characteristic. 
It proposes LiGNN, a hardware-based locality-aware dropout and merge solution that can accelerate GNN training. 
Experimental results show that under classic 0.5 droprate, LiGNN achieves 1.48$\sim$3.02$\times$ speedup, reduces DRAM accesses by 34$\sim$55\% and DRAM row activation by 59$\sim$82\%, all without losing accuracy.

%% file: tex/bio.tex
\section{Biography Section}

\begin{IEEEbiography}[{\includegraphics[width=1in,height=1.25in,clip,keepaspectratio]{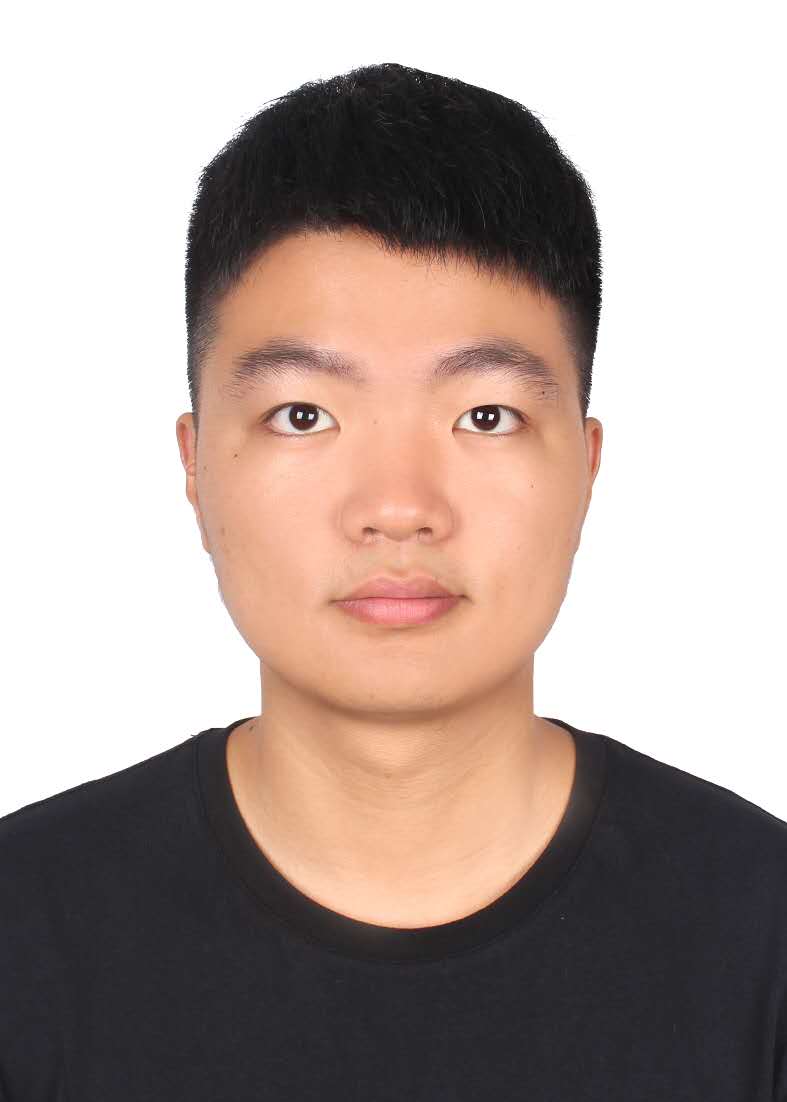}}]{Gongjian Sun} received the B.S. degree from University of Chinese Academy of Sciences, Beijing, China in 2019. He is currently working toward the Ph.D. degree at Institute of Computing Technology, Chinese Academy of Sciences, Beijing, China. His current research interests include graph neural network acceleration and domain-specific acceleration. To date, he has published two research papers in international compute journals and conferences, including IEEE TC, DATE and so on.
\end{IEEEbiography}

\begin{IEEEbiography}[{\includegraphics[width=1in,height=1.25in,clip,keepaspectratio]{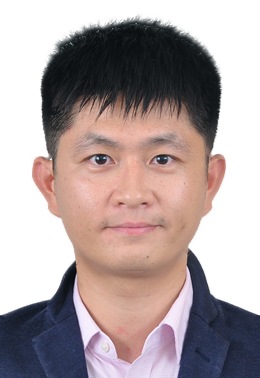}}]{Mingyu Yan} received the Ph.D. degree from University of Chinese Academy of Sciences, Beijing, China in 2020. He is currently an Associate Professor at Institute of Computing Technology, Chinese Academy of Sciences, Beijing, China. His current research interests is domain-specific hardware architecture for graph-based machine learning. To date, Dr. Yan has published over 20 research papers in top-tier journals and conferences, including the MICRO, HPCA, AAAI, IJCAI, DAC, ICCAD, PIEEE, IEEE TPDS, IEEE TC, IEEE TCAD, IEEE/CAA JAS, IEEE J-STSP, and so on. He has served as the TPC or ERC member for ISCA, HPCA, MICRO, and ICS.

\end{IEEEbiography}

\begin{IEEEbiography}[{\includegraphics[width=1in,height=1.25in,clip,keepaspectratio]{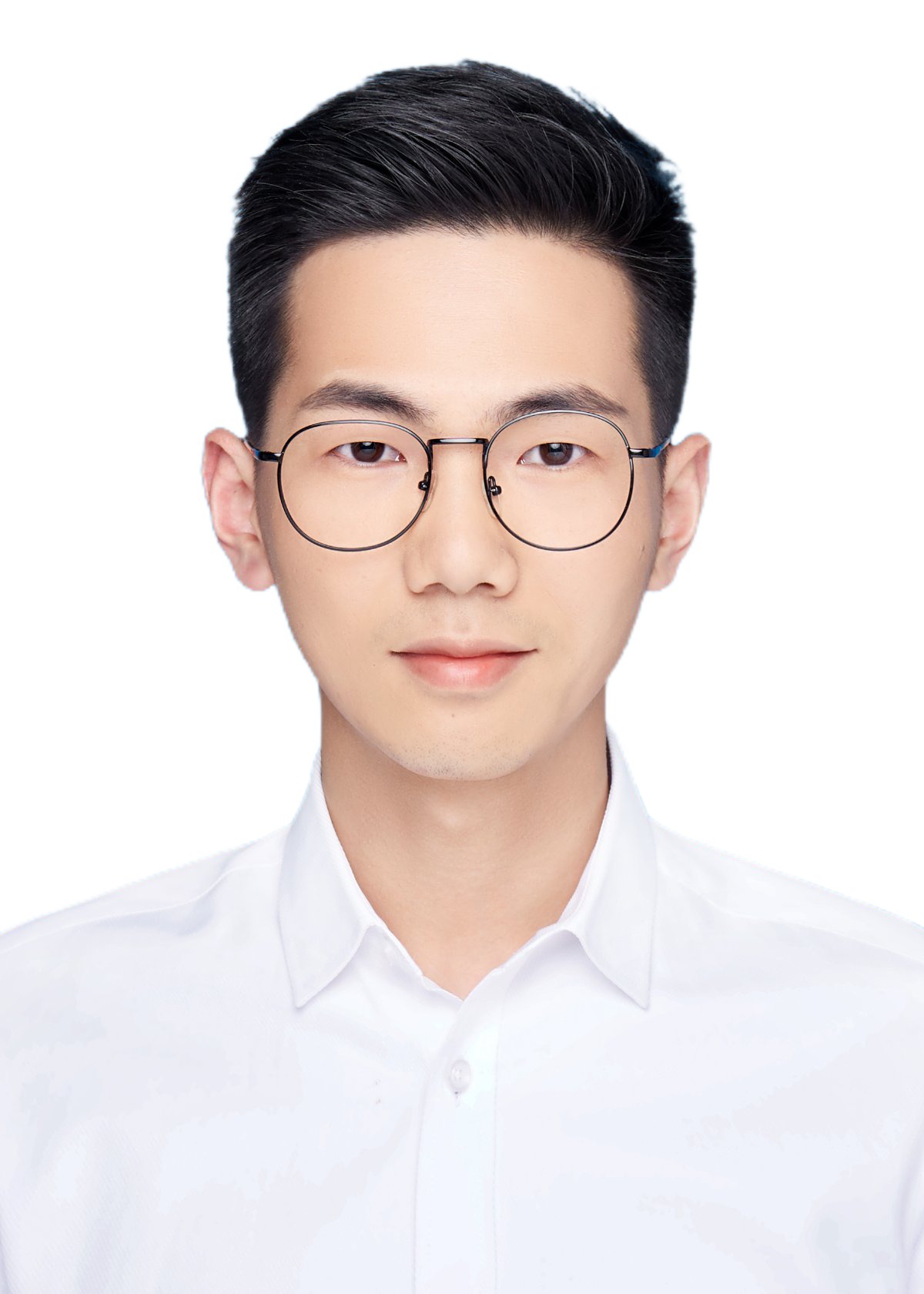}}]{Dengke Han} is currently a Ph.D. candidate at the Institute of Computing Technology, Chinese Academy of Sciences, Beijing, China. His research interests include graph-based hardware accelerators, algorithm performance analysis and optimization, and high-throughput computer architecture.
\end{IEEEbiography}

\begin{IEEEbiography}[{\includegraphics[width=1in,height=1.25in,clip,keepaspectratio]{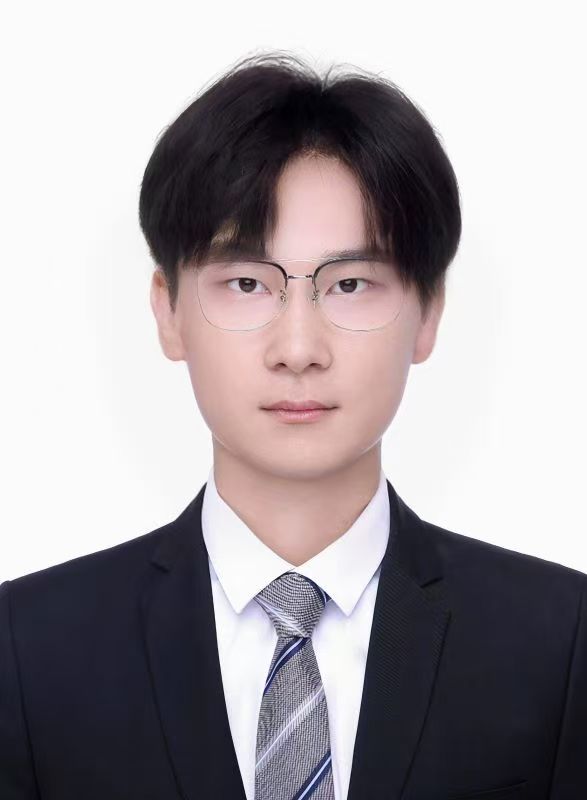}}]{Runzhen Xue} received his B.E. degree from Shandong University, Qingdao, China in 2021. He is currently a Ph.D. candidate at the Institute of Computing Technology, Chinese Academy of Sciences, Beijing, China. His research interests include the hardware accelerator, high-performance computer architecture, and processor design space exploration.
\end{IEEEbiography}

\begin{IEEEbiography}[{\includegraphics[width=1in,height=1.25in,clip,keepaspectratio]{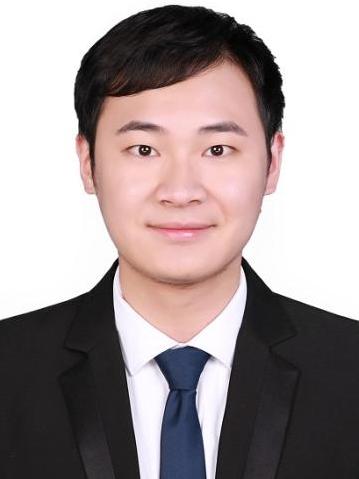}}]{Duo Wang} received the Ph.D. degree in computer architecture from Institute of Computing Technology, Chinese Academy of Sciences, Beijing, in 2024. He is currently a research associate in Institute of Computing Technology, Chinese Academy of Sciences, Beijing. His current research interests include processor design space exploration, high-performance computer architecture and software simulation.
\end{IEEEbiography}

\begin{IEEEbiography}[{\includegraphics[width=1in,height=1.25in,clip,keepaspectratio]{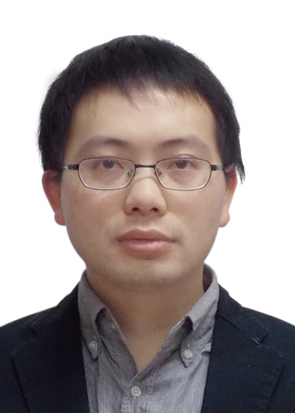}}]{Xiaochun Ye} received the Ph.D. degree in computer architecture from Institute of Computing Technology, Chinese Academy of Sciences, Beijing, in 2010. 
He is currently a Professor at Institute of Computing Technology, Chinese Academy of
Sciences, Beijing. 
His main research interest is domain-specific hardware architecture for graph-based machine learning and high-throughput computer architecture.
To date, Dr. Ye has published over 90 research papers in compute journals and conferences, the MICRO, HPCA, PACT, IPDPS, PIEEE, IEEE TC, IEEE TPDS, IEEE TACO, etc.

\end{IEEEbiography}

\begin{IEEEbiography}[{\includegraphics[width=1in,height=1.25in,clip,keepaspectratio]{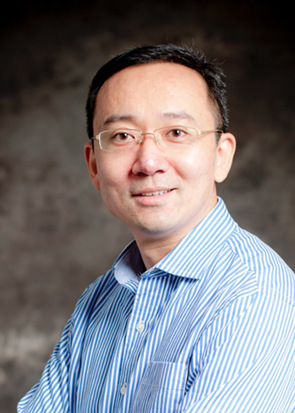}}]{Dongrui Fan} received the Ph.D. degree in computer architecture from Institute of Computing Technology, Chinese Academy of Sciences, Beijing, in 2005. 
He is currently a Professor and Ph.D. Supervisor at Institute of Computing Technology, Chinese Academy of Sciences, Beijing. 
His main research interests include high-throughput computer architecture, high-performance computer architecture, and low-power design.
To date, Dr. Fan has published over 140 research papers in compute journals and conferences, including the MICRO, HPCA, PPoPP, IJCAI, PACT, PIEEE, IEEE TC, IEEE TPDS, IEEE TCAD, IEEE TACO, IEEE Micro, and so on. He has been a member of the program committee of many important academic conferences, including MICRO, HPCA, etc.

\end{IEEEbiography}